\documentclass[letterpaper,twocolumn,amsmath,english,aps,prl,preprintnumbers,groupaddress,nofootinbib,tightenlines]{revtex4}
\usepackage[T1]{fontenc} \usepackage[latin9]{inputenc}
\usepackage{amsmath} \usepackage{varwidth} \usepackage{algorithm}
\usepackage{algorithmicx} \usepackage{algpseudocode}
\usepackage{graphicx} \usepackage{relsize}  

\usepackage{babel}
\begin{document}

\title{An {\em N}-Body Solution to the Problem of Fock Exchange}
\author{Matt Challacombe} 
\preprint{\texttt{LA-UR-14-20354}}

\email{matt.challacombe@freeon.org}
\thanks{corresponding author}
\homepage{http://www.freeon.org}
\author{Nicolas Bock}

\affiliation{Theoretical Division, Los Alamos National Laboratory}

\begin{abstract}
  We report an {\em N}-Body approach to computing the Fock exchange matrix
  with and without permutational symmetry.  The method achieves an
  $O(N {\rm lg} N)$ computational complexity through an embedded
  metric-query, allowing hierarchical application of direct SCF
  criteria. The advantages of permutational symmetry are found to be
  4-fold for small systems, but decreasing with increasing system size
  and/or more permissive neglect criteria.  This work sets the stage
  for: (1) the introduction of range queries in multi-level multipole
  schemes for rank reduction, and (2) recursive task parallelism.
\end{abstract}
\maketitle

\section{introduction}

For physical problems that find compact representation supported by
fast transforms, such as the fast wavelet transform and the fast
Fourier transform, recursion and reduced complexity are intrinsic.
For problems that do not find compact representation, there may be
fast {\em N}-Body solutions.  {\em N}-Body solvers combine recursive
subdivision, elements of database theory that include efficient
metric- and range-queries, as well as multi-level approaches to rank
reduction and culling.  In addition to a reduced computational
complexity, these database elements enable the exploitation of
temporal and spatial locality in high performance implementations,
making the {\em N}-Body programming model one of the most successful
in large scale scientific simulation
\cite{Warrena,Taiji,Fukushige1996,Warren1997,Kawai1999,Makino2003,Hamada2009,Hamada2010,Ishiyama:2012:PAN:2388996.2389003}. So
far, it has been possible to develop reduced complexity, {\em N}-Body
solutions for many aspects of conventional quantum chemical
self consistent field theory \cite{szabo_ostlund}, including the
Hartree problem
\cite{White:1996:CFMM,challacombe:4685,Challacombe:1997:QCTC,Challacombe:Review},
exchange correlation cubature \cite{Challacombe:2000:HiCu}, inverse
factorization and spectral projection based on the {\tt SpAMM}
algorithm \cite{Challacombe2010c,Bock2013a}.

Beyond the recursive multi-wavelet work of Yanai {\em et al.}~\cite{Yanai2004}, {\em N}-Body solutions to Fock exchange in the
conventional, Gaussian Type Atomic Orbital (GTAO) representation
remains an open and important problem. It's important because the Fock
exchange is a key ingredient in hybrid theories
\cite{becke:1372,becke:5648} that yield qualitatively better results
for many challenging problems (relative to ``pure'' DFT), from metal
oxides \cite{Perdew:1983:DFTFAIL,Sham:1985:DFTFAIL,Xiao:2011:DFTFAIL,
  PhysRevB.61.5194,PhysRevB.67.035403,Muscat2001397,
  wang2011electronic, wilson2012investigation}, to chemical reactions
\cite{Brittain:2009:DFTFAIL,Zhao:2011:DFTFAIL} battery materials
\cite{ramzan2011hybrid}, photovoltaic semiconductors
\cite{xiao2011accurate}, and in biochemistry
\cite{springerlink:10.1007/s00775-009-0469-9,
  rao2008performance,marianski2012comparison}.

Reduced complexity approaches to GTAO Fock exchange, such as ONX and
its variants
\cite{Schwegler1996,Schwegler1997,Schwegler1998,Ochsenfeld1998,Schwegler1999a,Schwegler1999,Schwegler,Schwegler2000a},
remain predicated on density matrix truncation (sparsification), preordered
skip-out lists and/or matrix range estimates, as well as list-of-lists
matrix structures such as BCSR \cite{Challacombe:2000:SpMM}.  These
preprocessing steps and their associated data structures greatly
complicate  aspects of the problem involving range-queries
and metric-queries necessary to implement multi-level methods for
rank reduction and culling \cite{Schwegler1999,Maurer2012}, and also
force a choice between ``integral driven'' and ``index driven'' schemes for
domain decomposition \cite{Takashima2002,Weber2006a}.  In this contribution, we
outline an {\em N}-body reformulation of Fock exchange matrix
construction that enables task decomposition in the recursive space of
tensor products, and which also provides an embedded framework for
metric- and range-queries.

\section{RECURSIVE QUANTUM CHEMISTRY}
We begin our development with recursive bisection of the indicial
space of GTAO basis functions:
\begin{equation}
  \left|\boldsymbol{ \mu}\right)^k =
  \left[
    \begin{array}{c} \left|\boldsymbol{ \mu}_0\right)^{k+1} \\[.1cm] \hline 
      \left|\boldsymbol{ \mu}_1\right)^{k+1}  
    \end{array}\right]
  =
  \left[
    \begin{array}{c} \phi_{\mu^L_0}     \\
      \phi_{\mu^{L+1}_0} \\
      \vdots             \\
      \phi_{\mu^R_0}     \\  \hline 
      \phi_{\mu^L_1}     \\
      \phi_{\mu^{L+1}_1} \\          
      \vdots             \\
      \phi_{\mu^R_1} 
    \end{array}\right]\; ,                 
\end{equation}
where $\left|\boldsymbol{ \mu}\right)^k$ denotes a vector (block) of functions
$\phi$ at depth $k$ with span $\mu \in \left[\mu^L,\mu^R \right]$.
This bisection can be accomplished in a variety of ways.  Here, we
consider bisection with ragged edges, rather than by powers of two, so
that the underlying atomic-orbital shell structure ({\it ie.}~{$p, sp,
  d, f\dots$) is preserved, greatly simplifying the associated
  computation of two-electron integrals.

  The fundamental telescoping quantities are matrix quadtrees,
  discussed in Refs.~[\onlinecite{Challacombe2010c,Bock2013a}] and
  references there in, and shell pairs, quadtrees that obtain
  recursively from shell-shell outerproducts:
  \begin{eqnarray}
    \left|\boldsymbol{ \mu \nu}\right)^{k}
    &=&\left(\begin{array}{cc}
        \left|\boldsymbol{\mu}_0 \boldsymbol{\nu}_0 \right)^{k+1} \; 
        & \left|\boldsymbol{\mu}_0 \boldsymbol{\nu}_1 \right)^{k+1} \\[0.15cm]
        \left|\boldsymbol{\mu}_1 \boldsymbol{\nu}_0 \right)^{k+1} \; 
        & \left|\boldsymbol{\mu}_1 \boldsymbol{\nu}_1 \right)^{k+1}
      \end{array}\right) \\[.25cm]
    &=&\left(\begin{array}{cc}
        \left|\boldsymbol{ \mu}_0\right)^{k+1}  \otimes  \left|\boldsymbol{ \nu}_0\right)^{k+1}  \quad
        & \left|\boldsymbol{ \mu}_0\right)^{k+1}  \otimes  \left|\boldsymbol{ \nu}_1\right)^{k+1} \\[0.15cm]
        \left|\boldsymbol{ \mu}_1\right)^{k+1} \otimes  \left|\boldsymbol{ \nu}_0\right)^{k+1} \quad
        &  \left|\boldsymbol{ \mu}_1\right)^{k+1}  \otimes  \left|\boldsymbol{ \nu}_1\right)^{k+1} \end{array}\right) \nonumber
  \end{eqnarray}
  In the large system limit, the complexity with respect to number of
  GTAO basis functions $N$ becomes ${\cal O}(N)$ due to
  non-overlapping functions,
  \begin{equation}
    \left|\boldsymbol{ \mu \nu}\right)^{k} =
    \begin{cases} 
      0 &  {\tt if} \; \left|\boldsymbol{ \mu}\right)^{k} \cap \left|\boldsymbol{ \nu}\right)^{k} = \emptyset \\
      \left|\boldsymbol{ \mu}\right)^{k} \otimes \left|\boldsymbol{
          \nu}\right)^{k} & \tt else
    \end{cases}
  \end{equation}
  where non-intersection obtains if all overlap integrals between
  shell-blocks are sufficiently small,
  \begin{multline}
    \left|\boldsymbol{ \mu}\right)^{k} \cap \left|\boldsymbol{
        \nu}\right)^{k} = \emptyset \quad {\tt if} \quad
    \left(\phi_\mu , \phi_\nu \right) < \tau_{\tt ovlp} \\ \forall \;
    \mu \in \left[\mu^L,\mu^R \right], \; \nu \in \left[\nu^L,\nu^R
    \right] \;,
  \end{multline}
  as determined by an overlap threshold $\tau_{\tt ovlp}$.

\section{EXCHANGE AS HEXTREE TRAVERSAL}
At the top level ($k=0$), the Fock exchange matrix can be written
simply with the less than orthodox bra-ket notation:
\begin{equation} {\bf K}^0 = - \frac{1}{2} \left(\boldsymbol{\mu}_0
    \boldsymbol{\nu}_0 \right|^{0} {\bf P}^0
  \left|\boldsymbol{\lambda}_0 \boldsymbol{\sigma}_0 \right)^{0} \;,
\end{equation}
which is useful shorthand for the tensor contraction
\begin{equation}
  K_{\mu \sigma}=-\frac{1}{2} \sum_{\nu,\lambda} P_{\nu,\lambda} \left(\mu \nu \right| \left. \lambda \sigma \right) \;,
\end{equation}
where $\left(\mu \nu \right| \left. \lambda \sigma \right)$ is the
standard two-electron integral over GTAO basis functions
\cite{szabo_ostlund}.

In the case of naive recursion, where permutational symmetry of the
two-electron integrals is unexploited, shell-pair quadtrees maintain
their relationship with sub-indices on recursion, permitting the
simplified notation: $\left|\boldsymbol{0 0}\right)^{k} \equiv
\left|\boldsymbol{ \mu_0 \nu_0}\right)^{k} $, $\left|\boldsymbol{0
    1}\right)^{k} \equiv \left|\boldsymbol{ \mu_0 \nu_1}\right)^{k} $
and so on.  Then, at all levels, sub-blocks of the Fock exchange matrix are
\begin{subequations}
  \begin{eqnarray}
    {\bf K}^k_{\boldsymbol{ 00 }} 
    &\leftarrow&  {\bf K}^k_{\boldsymbol{ 00 }}  - \left[  \left( \boldsymbol{ 00 } \right|^k {\bf P}^k_{\boldsymbol{ 00 }} \left| \boldsymbol{ 00 } \right)^k  
      +\left( \boldsymbol{ 00 } \right|^k {\bf P}^k_{\boldsymbol{01}} \left| \boldsymbol{10} \right)^k \right. \nonumber \\
    && \left.                               +\left( \boldsymbol{10} \right|^k {\bf P}^k_{\boldsymbol{10}} \left| \boldsymbol{ 00 } \right)^k 
      +\left( \boldsymbol{01} \right|^k {\bf P}^k_{\boldsymbol{11}} \left| \boldsymbol{10} \right)^k  \right] /2 \\
    {\bf K}^k_{\boldsymbol{01}} 
    &\leftarrow&  {\bf K}^k_{\boldsymbol{01}}  - \left[  \left( \boldsymbol{ 00 } \right|^k {\bf P}^k_{\boldsymbol{ 00 }} \left| \boldsymbol{01} \right)^k  
      +\left( \boldsymbol{01} \right|^k {\bf P}^k_{\boldsymbol{10}} \left| \boldsymbol{01} \right)^k \right. \nonumber \\
    && \left.                               +\left( \boldsymbol{ 00 } \right|^k {\bf P}^k_{\boldsymbol{01}} \left| \boldsymbol{11} \right)^k 
      +\left( \boldsymbol{01} \right|^k {\bf P}^k_{\boldsymbol{11}} \left| \boldsymbol{11} \right)^k  \right] /2\\
    {\bf K}^k_{\boldsymbol{10}} 
    &\leftarrow&  {\bf K}^k_{\boldsymbol{10}}  - \left[  \left( \boldsymbol{10} \right|^k {\bf P}^k_{\boldsymbol{ 00 }} \left| \boldsymbol{ 00 } \right)^k  
      +\left( \boldsymbol{11} \right|^k {\bf P}^k_{\boldsymbol{10}} \left| \boldsymbol{ 00 } \right)^k \right. \nonumber \\
    && \left.                               +\left( \boldsymbol{10} \right|^k {\bf P}^k_{\boldsymbol{01}} \left| \boldsymbol{10} \right)^k 
      +\left( \boldsymbol{11} \right|^k {\bf P}^k_{\boldsymbol{11}} \left| \boldsymbol{10} \right)^k  \right] /2  \\
    {\bf K}^k_{\boldsymbol{11}} 
    &\leftarrow&  {\bf K}^k_{\boldsymbol{11}}  - \left[  \left( \boldsymbol{10} \right|^k {\bf P}^k_{\boldsymbol{ 00 }} \left| \boldsymbol{01} \right)^k  
      +\left( \boldsymbol{11} \right|^k {\bf P}^k_{\boldsymbol{10}} \left| \boldsymbol{01} \right)^k \right. \nonumber \\
    && \left.                               +\left( \boldsymbol{10} \right|^k {\bf P}^k_{\boldsymbol{01}} \left| \boldsymbol{11} \right)^k 
      +\left( \boldsymbol{11} \right|^k {\bf P}^k_{\boldsymbol{11}} \left| \boldsymbol{11} \right)^k  \right] /2 
  \end{eqnarray}
\end{subequations}
equivalent to hextree traversal in the recursive task space 
of exchange tensor contraction.  An advantage of this construction is that, 
together with quadtrees that are informed at each level by the Frobenius norm 
$\Vert \cdot \Vert_{_F}$, the blocked Alml{\"o}f-Alhrichs criteria
\cite{JCC:JCC540030314,JCC:JCC540100111}
\begin{eqnarray}\label{AAcrit}
  \Vert\left(        {\boldsymbol \mu    } {\boldsymbol \nu    }\right| 
  \left.        {\boldsymbol \mu    } {\boldsymbol \nu    }\right)^k \Vert_{_F} \, \cdot \,
  \Vert {\bf P}^{k}_{ {\boldsymbol \nu         }{\boldsymbol \lambda}} \Vert_{_F}   \, \cdot \,
  \Vert\left(        {\boldsymbol \lambda    } {\boldsymbol \sigma    }\right| 
  \left.        {\boldsymbol \lambda    } {\boldsymbol \sigma    }\right)^k \Vert_{_F}
  \leq \mathlarger{\tau}_{2e} \; ,
\end{eqnarray}
can be carried out naturally via embedded metric-query, enabling
recursion termination when the bound is satisfied.  Because $\Vert
\cdot \Vert_{_F}$ is sub-multiplicative, this procedure is rigorously
equivalent to the standard direct SCF method, with the integral
threshold $\mathlarger{\tau}_{2e}$ retaining its conventional meaning.

The efficiency of this query in culling negligible integral
contributions is dependent on the numerical structure
of the the underlying data.  A simple solution to this problem involves ordering
shells with a locality preserving space filling curve, effectively
clustering elements of like magnitude as shown in Figs.~1 and 2
of Ref.~[\onlinecite{Bock2013a}].   Note that
methods relying on random permutation to make use of Cannon's
algorithm for the parallel multiplication of sparse matrices
\cite{Buluc:2008:SpMM,buluc2011parallel} destroy
these locality properties.  As with the {\tt SpAMM}
\cite{Challacombe2010c,Bock2013a} solver, truncation of the vector
space (sparsification) is not a prerequisite for achieving reduced
complexities, but well structured density matrices with decay are.

\section{RECURSION WITH SYMMETRY}

Extending the recursive approach outlined in the previous section to
the exploitation of 4-fold permutational symmetry is more involved
than in conventional schemes \cite{Ochsenfeld1998,Schwegler2000a}.
Naively, we wish to employ
\begin{subequations}
  \begin{eqnarray}
    {\bf K}^{k}_{ {\boldsymbol \mu    } {\boldsymbol \sigma }} \!\!\! &\leftarrow &\!\!\!  
    {\bf K}^{k}_{ {\boldsymbol \mu    } {\boldsymbol \sigma }} \!\! - 
    \left(        {\boldsymbol \mu    } {\boldsymbol \nu    }   \right|^k  
    {\bf P}^{k}_{ {\boldsymbol \nu    }{\boldsymbol \lambda}}  
    \left|        {\boldsymbol \lambda} {\boldsymbol \sigma } \right)^k /2 \\
    {\bf K}^{k}_{{\boldsymbol \nu    } {\boldsymbol \sigma }} \!\!\! &\leftarrow &\!\!\!  
    {\bf K}^{k}_{ {\boldsymbol \nu    } {\boldsymbol \sigma }} \!\! - 
    \left(        {\boldsymbol \nu    } {\boldsymbol \mu    }   \right|^k  
    {\bf P}^{k}_{ {\boldsymbol \mu    } {\boldsymbol \lambda}}  
    \left|        {\boldsymbol \lambda} {\boldsymbol \sigma } \right)^k /2 \\
    {\bf K}^{k}_{ {\boldsymbol \mu    } {\boldsymbol \lambda}} \!\!\! &\leftarrow &\!\!\!  
    {\bf K}^{k}_{ {\boldsymbol \mu    } {\boldsymbol \lambda}} \!\! - 
    \left(        {\boldsymbol \mu    } {\boldsymbol \nu    }   \right|^k  
    {\bf P}^{k}_{ {\boldsymbol \nu    } {\boldsymbol \sigma}}  
    \left|        {\boldsymbol \sigma } {\boldsymbol \lambda } \right)^k /2 \\
    {\bf K}^{k}_{ {\boldsymbol \nu    } {\boldsymbol \lambda}} \!\!\! &\leftarrow &\!\!\!  
    {\bf K}^{k}_{ {\boldsymbol \nu    } {\boldsymbol \lambda}} \!\! - 
    \left(        {\boldsymbol \nu    } {\boldsymbol \mu    }   \right|^k  
    {\bf P}^{k}_{ {\boldsymbol \mu    } {\boldsymbol \sigma}}  
    \left|        {\boldsymbol \sigma } {\boldsymbol \lambda } \right)^k /2 
  \end{eqnarray}\label{EqA}
\end{subequations}
which has the potential to yield speedups of up to {\tt 4x} in the
evaluation of two-electron integrals, but which does not reduce the cost of 
tensor contraction.  To avoid overcomputing however, the blockwise restriction
\begin{eqnarray}
  && \mu     \le \nu, \;\;  \lambda \le \sigma, \;\;  \mu+\nu(\nu-1)/2 \le \lambda+\sigma(\sigma-1)/2  \nonumber \\  
  && \forall  \; \mu \in \left[\mu^L,\mu^R \right], \quad \nu \in \left[\nu^L,\nu^R \right],  \\
  &&  \lambda \in \left[\lambda^L,\lambda^R \right],\quad {\rm and} \quad \sigma \in \left[\sigma^L,\sigma^R \right]. \nonumber 
\end{eqnarray}
must be observed at each level of recursion. Satisfying these
inequalities without carrying along an explosion of auxiliary source
and sink sub-matrices requires that we untether the strict 1-to-1
relationship between target symmetries and matrix sub-indices,
putting the transpose operation into play for shell-pairs and
matrices.  These complications are resolved by introducing an
intermediate level of recursion, where target symmetries are
determined and links to density and exchange matrix sub-blocks are
set, up to 4 of 16 possible links each. Including Eq.~(9), there are 8
conditions that exploit the full 4-fold symmetry of Fock exchange
under recursion (the remaining 7 are given in the Appendix), which
involve non-standard, ``across the bar'' (transpose) permutations as
well as additional factors of two, ({\em e.g.}~due to restricting the
density matrix above the diagonal.  In addition to cases of 4-fold
symmetry, there are many other intermediate and terminal cases where
there are fewer than 4 sub-matrices involved in recursion, for example
due to cases were blockwise symmetry operations do not yield a full
4-fold compliment, and also in cases where the density matrix is
sparse and sub-blocks are not available.  These later cases that fall
outside of Eqs.~(\ref{EqA})  \& (\ref{EqB}-\ref{EqH}) are referred to as
``sparse'' in the following. Also, the blockwise 
culling of negligible integral contributions
is carried out recursively,  as in Eq.~(\ref{AAcrit}), but instead using 
the maximum density matrix norms as they occur at each level.

\begin{figure}[]
  \centering
  \includegraphics[width=3.5in]{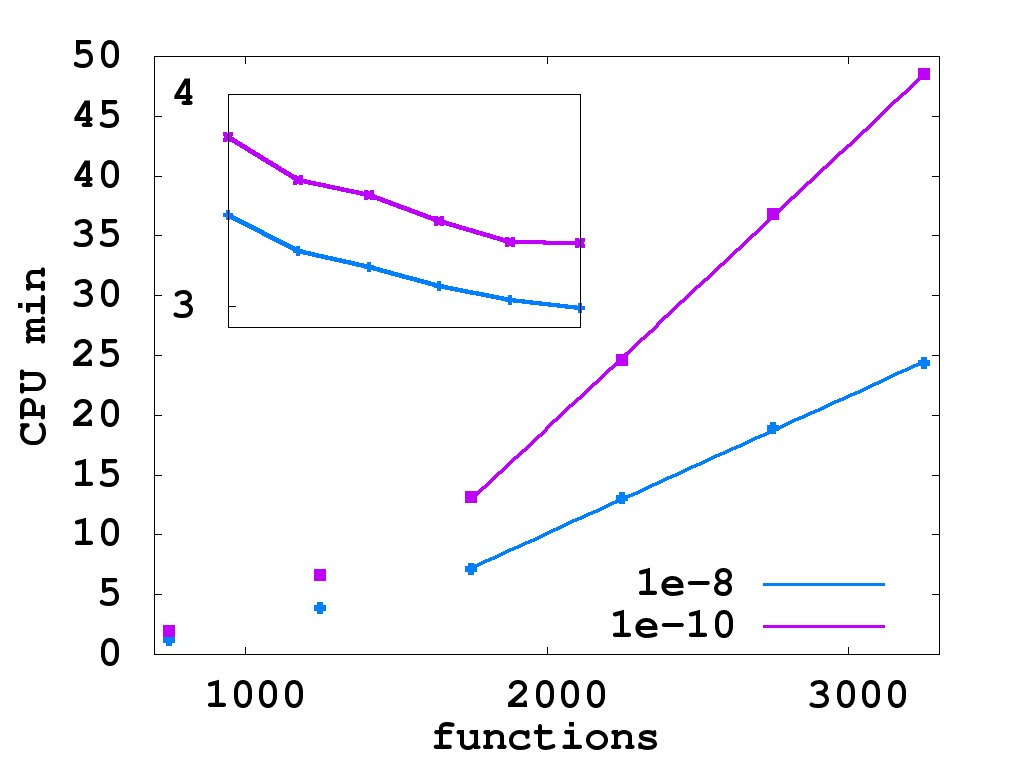}
  \caption{Onset of linear scaling for the symmetry enhanced {\em N}-Body Fock exchange solver  
           using two screening regimes: $(\mathlarger{\tau}_{\rm 2e},\mathlarger{\tau}_{\rm ovlp})=$ 
           ({\tt 1d-8},{\tt 1d-11}) and ({\tt 1d-10},{\tt 1d-13}).  
           The inset is the ratio of CPU times for  naive recursion relative to symmetry enhanced recursion.}
  \label{scaling1}
\end{figure}

\begin{figure}[]
  \centering
  \includegraphics[width=3.5in]{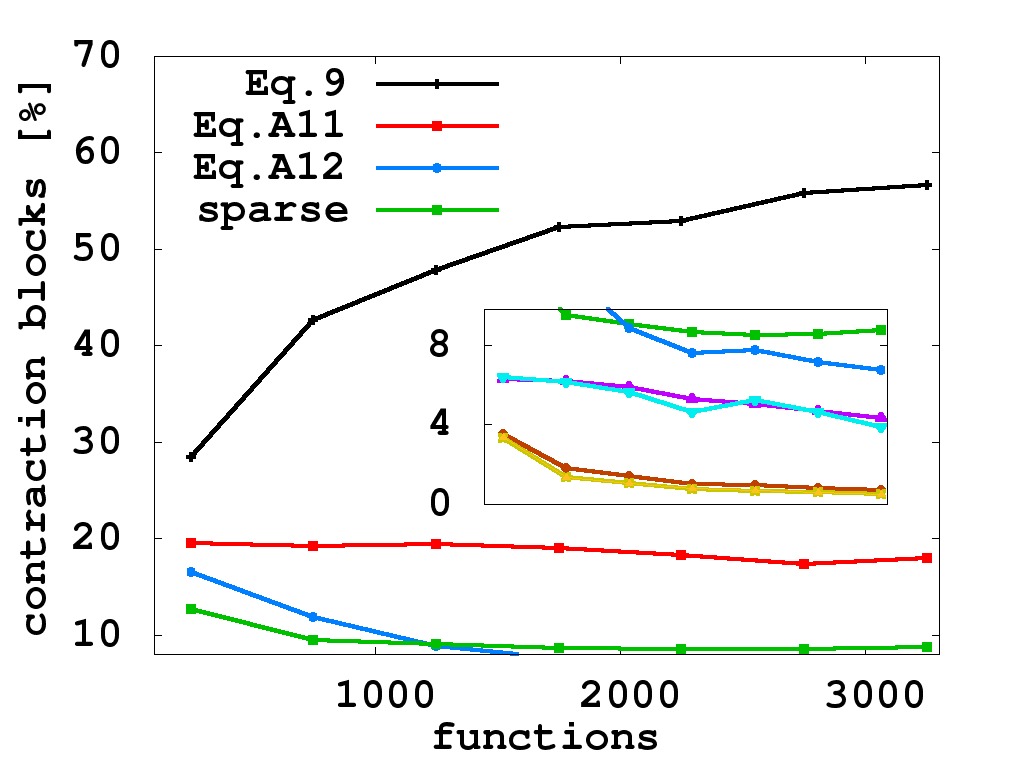}
  \caption{Percentage breakdown by contraction blocks, Eqs.~(\ref{EqA}) \& (\ref{EqB}-\ref{EqH}), including also 
    ``sparse'' blocks, in the symmetry enhanced {\em N}-Body Fock exchange solver in the tighter 
     ({\tt 1d-10},{\tt 1d-13}) regime.  Inset shows 
    blocks with fewer occurrences, corresponding to Eqs.~(\ref{EqC}-\ref{EqH}).}  
  \label{scaling2}
\end{figure}

\section{IMPLEMENTATION AND RESULTS}

Recursive construction of the Fock exchange matrix was implemented
with and without permutational symmetry in a development version of
{\tt FreeON} \cite{freeon}, using the Head-Gordon Pople algorithm for
two-electron integrals \cite{Head-Gordon1988}.  Non-sparse
contractions, involving full 4-fold matrix compliments, are carried
out in a single code block, different from conventional contraction
schemes in that there is no auxiliary ``batch'' dimension available
for vector optimization.  In the current implementation, recursion
extends to blocks with dimension {\tt 10x10} or smaller.  All
calculations were carried out on an Intel Xeon CPU {\tt E5-2687W} @
3.10GHz using v.~{\tt 14.0.1} of the Intel Fortran and C compilers.

Results are shown in Fig.~\ref{scaling1} and \ref{scaling2} for the
standard sequence of water droplets corresponding to STP
conditions,~(H$_2$O)$_{n}$ with $n=10,30,50,70,90,110, 130$, and
density matrices tightly converged using an atom blocked threshold
$\mathlarger{\tau}_{\rm mtrx}=${\tt 1d-6} \cite{Challacombe:2000:SpMM}
at the B3LYP/6-31G** level of theory.  All systems were ordered
using the Hilbert-curve to maximize locality as shown in Figs.~1 and 2
of Ref.~[\onlinecite{Bock2013a}].  In the largest droplet,
(H$_2$O)$_{130}$, ragged bisection with $\sim${\tt 10x10} blocks
yields a depth $k=10$, comprising $\sim${\tt 6d+5} nodes in the
shell-pair quadtree.

In Fig.~1, an early approach to linear scaling is shown for two values of the 
integral screening parameter $\mathlarger{\tau}_{\rm 2e}$, together with 
the relative performance of the naive vs symmetry enhanced methods as inset.  In the case of permutational symmetry, 
it is found that the cost of integral evaluation and contraction tend towards 1:1.  
Figure~2 shows the \% occurrence by symmetry block, Eqs.~(\ref{EqA}) \& (\ref{EqB}-\ref{EqH}), 
for the symmetry enhanced method.  The dominant occurrences correspond to Eq.~(\ref{EqA}) and 
(\ref{EqB}) at about 60\% and 20\% respectively, followed by the sparse case at 10\%. 
Because this recursive task space corresponds to a hextree, there
are ample opportunities for 
Certainly, the task parallel features of {\tt OpenMP 3.0}  \cite{5361951}
are ideally suited to this formulation, and middleware for distributed memory task parallelism, 
such as {\tt charm++} \cite{Kale93charm++:a},  
are becoming increasing powerful and simple to use.  
Also, the cost of integral contraction with and without symmetry is about 2:3, reflecting loop
overheads and higher levels of optimization that favor the combined 4-fold contraction blocks.
\section{Summary}

We've presented a novel, $N$-Body formulation for the naive and symmetry enhanced construction 
of the Fock exchange matrix, which achieves a reduced, linear scaling complexity for
density matrices with decay.  The method does not employ any of the matrix truncation (sparsification),
integral skipout lists, matrix range estimation or list-of-lists data structures employed by 
conventional methods.   Also, the  method may enable
transcendence of conventional  ``index driven'' vs ``integral'' driven paradigms for parallel Fock matrix 
construction \cite{Takashima2002,Weber2006a}, through decomposition in the recursive
task space of the tensor contraction.  Even though this task space is sparse and irregular due to 
culling,  because it is higher dimensional, corresponding to hextree traversal, it offers ample opportunity for 
parallelism through well developed middleware such as {\tt OpenMP 3.0}  \cite{5361951} in 
shared memory environments and {\tt charm++} \cite{Kale93charm++:a} in distributed environments. 

With this contribution, all minimally essential components of reduced complexity electronic structure theory 
at the self consistent field level of theory have been reformulated as {\em N}-Body solvers.   
It remains to be seen how tightly these solvers can be integrated  using a common infrastructure and programming model.  
Finally, it also remains to be seen if the ability to embed range- and metric-queries into this structure 
can be exploited to achieve true multi-level rank reduction for computation of the Fock exchange matrix. 

\begin{acknowledgments}
This work was supported by the U. S. Department of Energy under Contract No. DE-AC52-06NA25396 and LDRD-ER grant 20110230ER.  
The authors acknowledge the stimulating environment of the Ten-Bar Caf{\'e} and expert administrations of the barista.
\end{acknowledgments}


\appendix*
\section{Appendix}\label{AppendixA}
Additional symmetry blocks:
\begin{subequations}
  \begin{eqnarray}
    {\bf K}^{k}_{ {\boldsymbol \mu    } {\boldsymbol \sigma }} \!\!\! &\leftarrow &\!\!\!  
    {\bf K}^{k}_{ {\boldsymbol \mu    } {\boldsymbol \sigma }} \!\! - 
    \left(        {\boldsymbol \mu    } {\boldsymbol \nu    }   \right|^k  
    {\bf P}^{k}_{ {\boldsymbol \nu    }{\boldsymbol \lambda}}  
    \left|        {\boldsymbol \lambda} {\boldsymbol \sigma } \right)^k /2 \\
    {\bf K}^{k}_{{\boldsymbol \nu    } {\boldsymbol \sigma }} \!\!\! &\leftarrow &\!\!\!  
    {\bf K}^{k}_{ {\boldsymbol \nu    } {\boldsymbol \sigma }} \!\! - 
    \left(        {\boldsymbol \nu    } {\boldsymbol \mu    }   \right|^k  
    {\bf P}^{k}_{ {\boldsymbol \mu    } {\boldsymbol \lambda}}  
    \left|        {\boldsymbol \lambda} {\boldsymbol \sigma } \right)^k /2 \\
    {\bf K}^{k}_{ {\boldsymbol \lambda    } {\boldsymbol \mu}} \!\!\! &\leftarrow &\!\!\!  
    {\bf K}^{k}_{ {\boldsymbol \lambda    } {\boldsymbol \mu}} \!\! - 
    \left(        {\boldsymbol \lambda    } {\boldsymbol \sigma    }   \right|^k  
    {\bf P}^{k}_{ {\boldsymbol \sigma    } {\boldsymbol \nu}}  
    \left|        {\boldsymbol \nu } {\boldsymbol \mu } \right)^k /2 \\
    {\bf K}^{k}_{ {\boldsymbol \nu    } {\boldsymbol \lambda}} \!\!\! &\leftarrow &\!\!\!  
    {\bf K}^{k}_{ {\boldsymbol \nu    } {\boldsymbol \lambda}} \!\! - 
    \left(        {\boldsymbol \nu    } {\boldsymbol \mu    }   \right|^k  
    {\bf P}^{k}_{ {\boldsymbol \mu    } {\boldsymbol \sigma}}  
    \left|        {\boldsymbol \sigma } {\boldsymbol \lambda } \right)^k /2 
  \end{eqnarray}\label{EqB}
\end{subequations}
\begin{subequations}
  \begin{eqnarray}
    {\bf K}^{k}_{ {\boldsymbol \mu    } {\boldsymbol \sigma }} \!\!\! &\leftarrow &\!\!\!  
    {\bf K}^{k}_{ {\boldsymbol \mu    } {\boldsymbol \sigma }} \!\! - 
    \left(        {\boldsymbol \mu    } {\boldsymbol \nu    }   \right|^k  
    {\bf P}^{k}_{ {\boldsymbol \nu    }{\boldsymbol \lambda}}  
    \left|        {\boldsymbol \lambda} {\boldsymbol \sigma } \right)^k /2 \\
    {\bf K}^{k}_{ {\boldsymbol \nu    } {\boldsymbol \lambda}} \!\!\! &\leftarrow &\!\!\!  
    {\bf K}^{k}_{ {\boldsymbol \nu    } {\boldsymbol \lambda}} \!\! - 
    \left(        {\boldsymbol \nu    } {\boldsymbol \mu    }   \right|^k  
    {\bf P}^{k}_{ {\boldsymbol \mu    } {\boldsymbol \sigma}}  
    \left|        {\boldsymbol \sigma } {\boldsymbol \lambda } \right)^k /2 \\
    {\bf K}^{k}_{ {\boldsymbol \mu    } {\boldsymbol \lambda}} \!\!\! &\leftarrow &\!\!\!  
    {\bf K}^{k}_{ {\boldsymbol \mu    } {\boldsymbol \lambda}} \!\! - 
    \left(        {\boldsymbol \mu    } {\boldsymbol \nu    }   \right|^k  
    {\bf P}^{k}_{ {\boldsymbol \nu    } {\boldsymbol \sigma}}  
    \left|        {\boldsymbol \sigma } {\boldsymbol \lambda } \right)^k /2 \\
    {\bf K}^{k}_{ {\boldsymbol \lambda    } {\boldsymbol \nu}} \!\!\! &\leftarrow &\!\!\!  
    {\bf K}^{k}_{ {\boldsymbol \lambda    } {\boldsymbol \nu}} \!\! - 
    \left(        {\boldsymbol \lambda    } {\boldsymbol \sigma    }   \right|^k  
    {\bf P}^{k}_{ {\boldsymbol \sigma    } {\boldsymbol \mu}}  
    \left|        {\boldsymbol \mu } {\boldsymbol \nu } \right)^k /2 
  \end{eqnarray}\label{EqC}
\end{subequations}
\begin{subequations}
  \begin{eqnarray}
    {\bf K}^{k}_{ {\boldsymbol \mu    } {\boldsymbol \sigma }} \!\!\! &\leftarrow &\!\!\!  
    {\bf K}^{k}_{ {\boldsymbol \mu    } {\boldsymbol \sigma }} \!\! - 
    \left(        {\boldsymbol \mu    } {\boldsymbol \nu    }   \right|^k  
    {\bf P}^{k}_{ {\boldsymbol \nu    }{\boldsymbol \lambda}}  
    \left|        {\boldsymbol \lambda} {\boldsymbol \sigma } \right)^k /2 \\
    {\bf K}^{k}_{ {\boldsymbol \nu    } {\boldsymbol \sigma}} \!\!\! &\leftarrow &\!\!\!  
    {\bf K}^{k}_{ {\boldsymbol \nu    } {\boldsymbol \sigma}} \!\! - 
    \left(        {\boldsymbol \nu    } {\boldsymbol \mu    }   \right|^k  
    {\bf P}^{k}_{ {\boldsymbol \mu    } {\boldsymbol \lambda}}  
    \left|        {\boldsymbol \lambda } {\boldsymbol \sigma } \right)^k /2 \\
    {\bf K}^{k}_{ {\boldsymbol \mu    } {\boldsymbol \lambda}} \!\!\! &\leftarrow &\!\!\!  
    {\bf K}^{k}_{ {\boldsymbol \mu    } {\boldsymbol \lambda}} \!\! - 
    \left(        {\boldsymbol \mu    } {\boldsymbol \nu    }   \right|^k  
    {\bf P}^{k}_{ {\boldsymbol \nu    } {\boldsymbol \sigma}}  
    \left|        {\boldsymbol \sigma } {\boldsymbol \lambda } \right)^k /2 \\
    {\bf K}^{k}_{ {\boldsymbol \lambda    } {\boldsymbol \nu}} \!\!\! &\leftarrow &\!\!\!  
    {\bf K}^{k}_{ {\boldsymbol \lambda    } {\boldsymbol \nu}} \!\! - 
    \left(        {\boldsymbol \lambda    } {\boldsymbol \sigma  } \right|^k  
    {\bf P}^{k}_{ {\boldsymbol \sigma    } {\boldsymbol \mu}}  
    \left|        {\boldsymbol \mu } {\boldsymbol \nu } \right)^k /2 
  \end{eqnarray}\label{EqD}
\end{subequations}
\begin{subequations}
  \begin{eqnarray}
    {\bf K}^{k}_{ {\boldsymbol \mu    } {\boldsymbol \sigma }} \!\!\! &\leftarrow &\!\!\!  
    {\bf K}^{k}_{ {\boldsymbol \mu    } {\boldsymbol \sigma }} \!\! - 
    \left(        {\boldsymbol \mu    } {\boldsymbol \nu    }   \right|^k  
    {\bf P}^{k}_{ {\boldsymbol \nu    }{\boldsymbol \lambda}}  
    \left|        {\boldsymbol \lambda} {\boldsymbol \sigma } \right)^k /2 \\
    {\bf K}^{k}_{ {\boldsymbol \nu    } {\boldsymbol \sigma}} \!\!\! &\leftarrow &\!\!\!  
    {\bf K}^{k}_{ {\boldsymbol \nu    } {\boldsymbol \sigma}} \!\! - 
    \left(        {\boldsymbol \nu    } {\boldsymbol \mu    }   \right|^k  
    {\bf P}^{k}_{ {\boldsymbol \mu    } {\boldsymbol \lambda}}  
    \left|        {\boldsymbol \lambda } {\boldsymbol \sigma } \right)^k /2 \\
    {\bf K}^{k}_{ {\boldsymbol \lambda    } {\boldsymbol \mu}} \!\!\! &\leftarrow &\!\!\!  
    {\bf K}^{k}_{ {\boldsymbol \lambda    } {\boldsymbol \mu}} \!\! - 
    \left(        {\boldsymbol \lambda    } {\boldsymbol \sigma    }   \right|^k  
    {\bf P}^{k}_{ {\boldsymbol \sigma    } {\boldsymbol \nu}}  
    \left|        {\boldsymbol \nu } {\boldsymbol \mu } \right)^k /2 \\
    {\bf K}^{k}_{ {\boldsymbol \lambda    } {\boldsymbol \nu}} \!\!\! &\leftarrow &\!\!\!  
    {\bf K}^{k}_{ {\boldsymbol \lambda    } {\boldsymbol \nu}} \!\! - 
    \left(        {\boldsymbol \lambda    } {\boldsymbol \sigma  } \right|^k  
    {\bf P}^{k}_{ {\boldsymbol \sigma    } {\boldsymbol \mu}}  
    \left|        {\boldsymbol \mu } {\boldsymbol \nu } \right)^k /2 
  \end{eqnarray}\label{EqE}
\end{subequations}
\begin{subequations}
  \begin{eqnarray}
    {\bf K}^{k}_{ {\boldsymbol \mu    } {\boldsymbol \sigma }} \!\!\! &\leftarrow &\!\!\!  
    {\bf K}^{k}_{ {\boldsymbol \mu    } {\boldsymbol \sigma }} \!\! - 
    \left(        {\boldsymbol \mu    } {\boldsymbol \nu    }   \right|^k  
    {\bf P}^{k}_{ {\boldsymbol \nu    }{\boldsymbol \nu}}  
    \left|        {\boldsymbol \nu} {\boldsymbol \sigma } \right)^k /2 \\
    {\bf K}^{k}_{{\boldsymbol \nu    } {\boldsymbol \sigma }} \!\!\! &\leftarrow &\!\!\!  
    {\bf K}^{k}_{ {\boldsymbol \nu    } {\boldsymbol \sigma }} \!\! - 
    \left(        {\boldsymbol \nu    } {\boldsymbol \mu    }   \right|^k  
    {\bf P}^{k}_{ {\boldsymbol \mu    } {\boldsymbol \nu}}  
    \left|        {\boldsymbol \nu} {\boldsymbol \sigma } \right)^k /2 \\
    {\bf K}^{k}_{ {\boldsymbol \mu    } {\boldsymbol \nu}} \!\!\! &\leftarrow &\!\!\!  
    {\bf K}^{k}_{ {\boldsymbol \mu    } {\boldsymbol \nu}} \!\! - 
    \left(        {\boldsymbol \mu    } {\boldsymbol \nu    }   \right|^k  
    {\bf P}^{k}_{ {\boldsymbol \nu    } {\boldsymbol \sigma}}  
    \left|        {\boldsymbol \sigma } {\boldsymbol \nu } \right)^k /2 \\
    {\bf K}^{k}_{ {\boldsymbol \nu    } {\boldsymbol \nu}} \!\!\! &\leftarrow &\!\!\!  
    {\bf K}^{k}_{ {\boldsymbol \nu    } {\boldsymbol \nu}} \!\! -  \left[
      \left(        {\boldsymbol \nu    } {\boldsymbol \sigma    }   \right|^k  
      {\bf P}^{k}_{ {\boldsymbol \sigma    } {\boldsymbol \mu}}  
      \left|        {\boldsymbol \mu } {\boldsymbol \nu } \right)^k  \right. \\
    &&\left. \qquad + 
      \left(        {\boldsymbol \nu    } {\boldsymbol \mu    }   \right|^k  
      {\bf P}^{k}_{ {\boldsymbol \mu    } {\boldsymbol \sigma}}  
      \left|        {\boldsymbol \sigma } {\boldsymbol \nu } \right)^k \right] /2 \nonumber
  \end{eqnarray}\label{EqF}
\end{subequations}
\begin{subequations}
  \begin{eqnarray}
    {\bf K}^{k}_{ {\boldsymbol \mu    } {\boldsymbol \nu }} \!\!\! &\leftarrow &\!\!\!  
    {\bf K}^{k}_{ {\boldsymbol \mu    } {\boldsymbol \nu }} \!\! - 
    \left(        {\boldsymbol \mu    } {\boldsymbol \nu    }   \right|^k  
    {\bf P}^{k}_{ {\boldsymbol \nu    }{\boldsymbol \nu}}  
    \left|        {\boldsymbol \nu} {\boldsymbol \nu } \right)^k /2 \\
    {\bf K}^{k}_{ {\boldsymbol \nu    } {\boldsymbol \nu}} \!\!\! &\leftarrow &\!\!\!  
    {\bf K}^{k}_{ {\boldsymbol \nu    } {\boldsymbol \nu}} \!\! -  \left[
      \left(        {\boldsymbol \nu    } {\boldsymbol \nu    }   \right|^k  
      {\bf P}^{k}_{ {\boldsymbol \nu    } {\boldsymbol \mu}}  
      \left|        {\boldsymbol \mu } {\boldsymbol \nu } \right)^k  \right. \\
    &&\left. \qquad + 
      \left(        {\boldsymbol \nu    } {\boldsymbol \mu    }   \right|^k  
      {\bf P}^{k}_{ {\boldsymbol \mu    } {\boldsymbol \nu}}  
      \left|        {\boldsymbol \nu } {\boldsymbol \nu } \right)^k \right] /2 \nonumber \\
    {\bf K}^{k}_{ {\boldsymbol \mu    } {\boldsymbol \lambda}} \!\!\! &\leftarrow &\!\!\!  
    {\bf K}^{k}_{ {\boldsymbol \mu    } {\boldsymbol \nu}} \!\! - 
    \left(        {\boldsymbol \mu    } {\boldsymbol \nu    }   \right|^k  
    {\bf P}^{k}_{ {\boldsymbol \nu    } {\boldsymbol \nu}}  
    \left|        {\boldsymbol \nu } {\boldsymbol \lambda } \right)^k /2 \\
    {\bf K}^{k}_{ {\boldsymbol \lambda    } {\boldsymbol \nu}} \!\!\! &\leftarrow &\!\!\!  
    {\bf K}^{k}_{ {\boldsymbol \lambda    } {\boldsymbol \nu}} \!\! - 
    \left(        {\boldsymbol \lambda    } {\boldsymbol \nu    }   \right|^k  
    {\bf P}^{k}_{ {\boldsymbol \nu    } {\boldsymbol \mu}}  
    \left|        {\boldsymbol \mu } {\boldsymbol \nu } \right)^k /2 
  \end{eqnarray} \label{EqF}
\end{subequations}
\begin{subequations}
  \begin{eqnarray}
    {\bf K}^{k}_{ {\boldsymbol \mu    } {\boldsymbol \sigma }} \!\!\! &\leftarrow &\!\!\!  
    {\bf K}^{k}_{ {\boldsymbol \mu    } {\boldsymbol \sigma }} \!\! - 
    \left(        {\boldsymbol \mu    } {\boldsymbol \nu    }   \right|^k  
    {\bf P}^{k}_{ {\boldsymbol \nu    }{\boldsymbol \mu}}  
    \left|        {\boldsymbol \mu} {\boldsymbol \sigma } \right)^k /2 \\
    {\bf K}^{k}_{{\boldsymbol \nu    } {\boldsymbol \sigma }} \!\!\! &\leftarrow &\!\!\!  
    {\bf K}^{k}_{ {\boldsymbol \nu    } {\boldsymbol \sigma }} \!\! - 
    \left(        {\boldsymbol \nu    } {\boldsymbol \mu    }   \right|^k  
    {\bf P}^{k}_{ {\boldsymbol \mu    } {\boldsymbol \mu}}  
    \left|        {\boldsymbol \mu} {\boldsymbol \sigma } \right)^k /2 \\
    {\bf K}^{k}_{ {\boldsymbol \mu    } {\boldsymbol \mu}} \!\!\! &\leftarrow &\!\!\!  
    {\bf K}^{k}_{ {\boldsymbol \mu    } {\boldsymbol \mu}} \!\! - 
    \left[ \left(        {\boldsymbol \mu    } {\boldsymbol \nu    }   \right|^k  
      {\bf P}^{k}_{ {\boldsymbol \nu    } {\boldsymbol \sigma}}  
      \left|        {\boldsymbol \sigma } {\boldsymbol \mu } \right)^k  \right. \\
    &&\left. \qquad + \left(        {\boldsymbol \mu    } {\boldsymbol \sigma    }   \right|^k  
      {\bf P}^{k}_{ {\boldsymbol \sigma    } {\boldsymbol \nu}}  
      \left|        {\boldsymbol \nu } {\boldsymbol \mu } \right)^k \right]/2\\
    {\bf K}^{k}_{ {\boldsymbol \nu    } {\boldsymbol \mu}} \!\!\! &\leftarrow &\!\!\!  
    {\bf K}^{k}_{ {\boldsymbol \nu    } {\boldsymbol \mu}} \!\! - 
    \left(        {\boldsymbol \nu    } {\boldsymbol \mu    }   \right|^k  
    {\bf P}^{k}_{ {\boldsymbol \mu    } {\boldsymbol \sigma}}  
    \left|        {\boldsymbol \sigma } {\boldsymbol \mu } \right)^k /2 
  \end{eqnarray}\label{EqH}
\end{subequations}

\end{document}